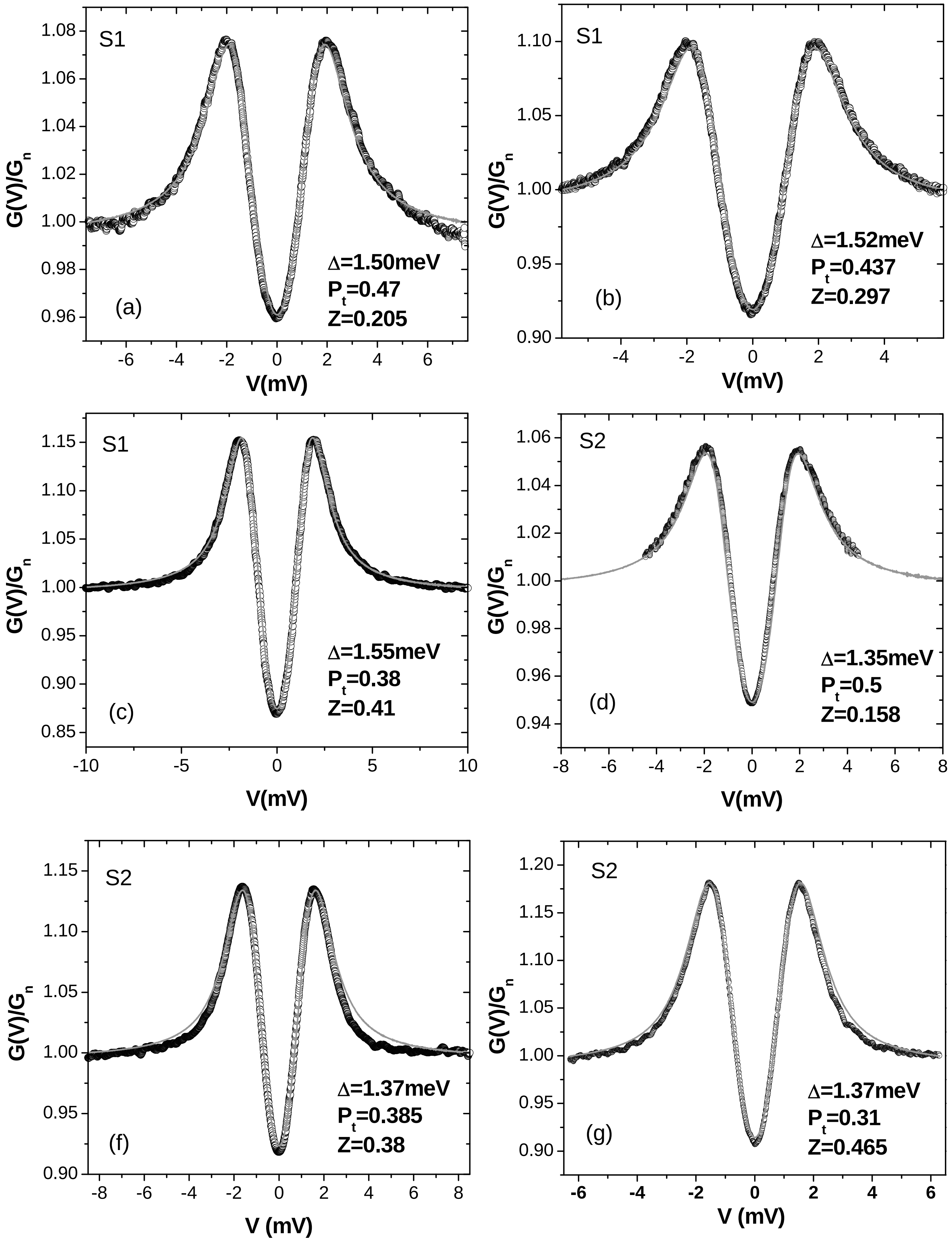

Figure 1

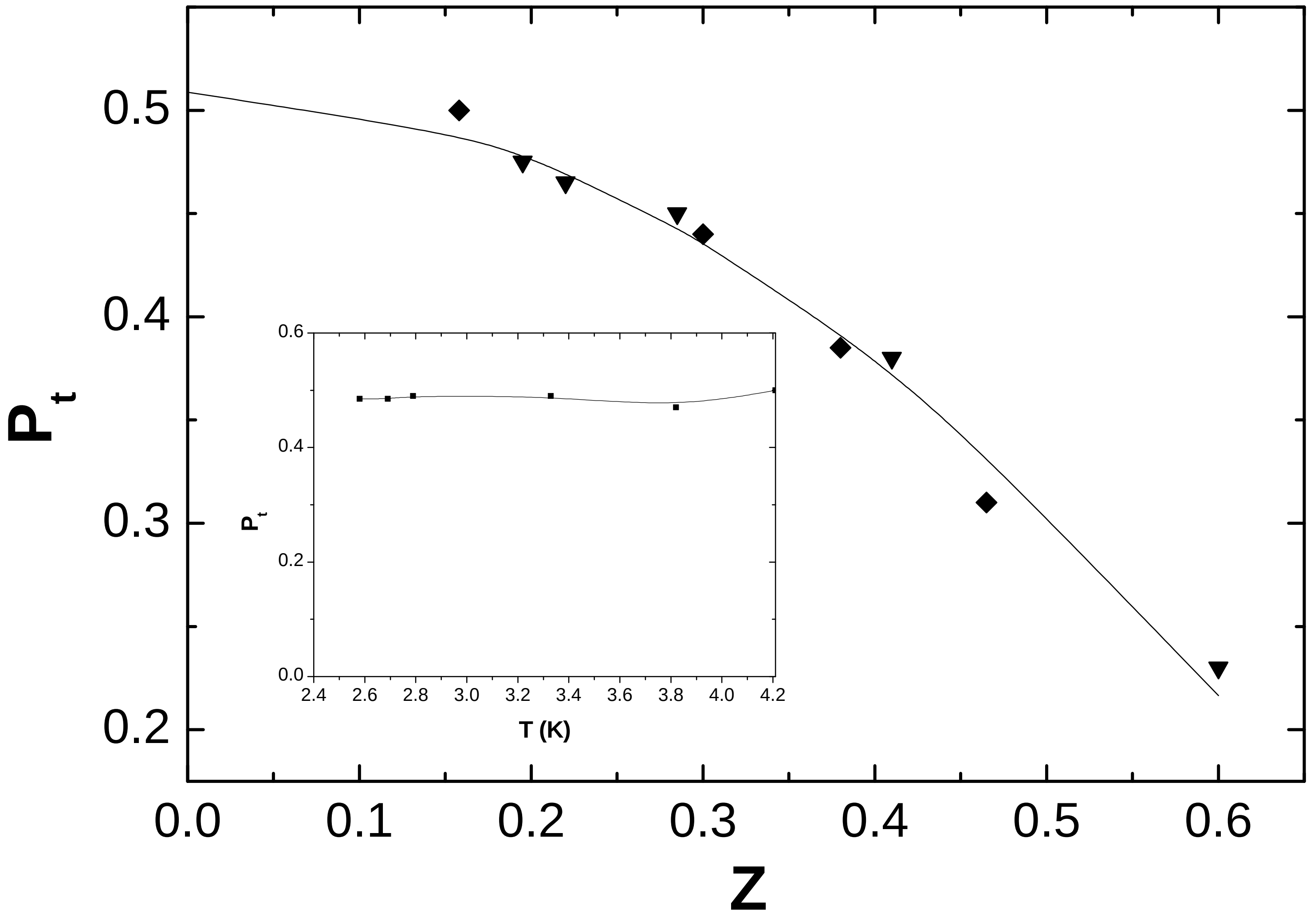

Figure 2

# Transport spin polarisation in $SrRuO_3$ measured through Point Contact Andreev reflection

P. Raychaudhuri[1,2], A. P. Mackenzie[1], J. W. Reiner[3] and M. R. Beasley[3]
[1]*Department of Physics and Astronomy, University of St. Andrews, North Haugh, St. Andrews, KY16 9SS, Scotland.*
[2]*School of Physics and Astronomy, University of Birmingham, Edgbaston, B15 2TT, UK.*
[3]*Edward L. Ginzton Laboratories, Stanford University, Stanford, California 94305*

**Abstract:** We report a study in which Andreev reflection using a Nb point contact is used to measure the transport spin polarisation of the 4d itinerant ferromagnet $SrRuO_3$. By performing the study in high quality thin films with residual resistivities less than 7μΩ-cm, we ensure that the study is done in the ballistic limit, a regime which is difficult to reach in oxide ferromagnets. The degree of transport spin polarisation that we find is comparable to that of the hole doped rare-earth manganites. We conclude that the large transport spin polarisation results mainly from a difference in the Fermi velocities between the majority and minority spin channels in this material.

Metallic oxide ferromagnets have been a field of great interest since the discovery of colossal magnetoresistance in doped rare-earth manganites[1,2]. Though the initial interest was triggered by the phenomenon of colossal magnetoresistance, it was soon realised that the large degree of spin polarisation observed in many of these oxides and the ability to grow these materials in epitaxial thin film form also made them potential candidates to explore novel forms of electronics where both the charge and the spin of the electrons could be used[3,4,5]. Towards this end, ferromagnetic tunnel junctions exhibiting both large positive and negative magnetoresistance using ferromagnetic oxides have already been fabricated[3]. Another field of great interest has been the effect of spin injection on the critical current of superconductors. It has been speculated that the low carrier densities and the existence of nodes in the superconducting gap in *d*-wave superconductors such as $YBa_2Cu_3O_7$ will cause an effective suppression of the critical current when spin polarised quasi-particles are injected in the superconductor. Experimental studies in this field have been primarily motivated by the fact that successful implementation of this phenomenon could result in fast switching devices with high gain. In this class of experiments, spin polarised carriers from an oxide ferromagnet are typically injected into a *d*-wave superconductor such as $YBa_2Cu_3O_7$ while the critical current of the superconductor is measured. While the ability to grow epitaxial superconducting layers on many of the perovskite oxide ferromagnets makes them attractive sources of spin polarised electrons for these experiments, quantitative interpretation of the data is crucially dependent on a prior knowledge of the degree of spin polarisation of the injected carriers across the ferromagnet/superconductor interface. This can, however, be difficult to calculate, since most perovskite oxide ferromagnets have complicated band structures with several bands crossing the Fermi surface, and further Fermi surface fragmentation often results from subtle crystalline distortions. It is therefore important to measure the degree of spin polarisation in these materials experimentally.

The spin polarisation in a ferromagnet is normally defined in terms of the difference in the density of states of spin up and spin down electrons at the Fermi level, namely, $P=(N_\uparrow(E_F)-N_\downarrow(E_F))/(N_\uparrow(E_F)+N_\downarrow(E_F))$. This quantity is not, however, very useful in transport experiments where a current is passed through an interface between the ferromagnet and another material. Here the polarisation of the injected current depends on the difference in the total flux of spin up and spin down electrons which depends both on the density of states and the Fermi velocities of the up and down electrons. In a ballistic point contact experiment, where the electron is allowed to flow from the ferromagnet to another metal/superconductor through an orifice smaller than the mean free path of the carriers (*l*), the net flux of of the carriers with spin $\sigma$ from a particular band i is given by[6] $\langle N_{i\sigma\mathbf{k}} v_{in\sigma\mathbf{k}} \rangle$, where **k** is the wave vector on the Fermi surface, **n** is the direction of the current flow and the average is taken over the Fermi surface. Explicitly carrying out the averaging, this quantity can be readily seen to be proportional to $S_{i\sigma n}$, the area of projection of the i-the band with spin σ on the interface plane. Summing over all the bands, the total flux of carriers with spin σ is given by $S_{\sigma n}$, the total area of projection of the bands with spin σ on the interface plane. Thus the spin polarisation of the injected current, commonly known as the transport spin polarisation, is given by the net difference of the area of projection of the up and down spin bands on the interface plane, namely, $P_t=((S_{\uparrow n}-S_{\downarrow n})/(S_{\uparrow n}+S_{\downarrow n}))$. $P_t$ can be either smaller or bigger than P and become identical to P only in an isotropic fermi surface and if the velocity of the up and down spins are equal. A knowledge of $P_t$ is directly relevant in interpreting results in experiments involving the transport of spin polarised carriers from a ferromagnet across an interface.

The 4*d* itinerant ferromagnet $SrRuO_3$ is an interesting material to explore from this point of view. $SrRuO_3$ stabilises in an orthorhombically distorted perovskite structure $a \approx b \approx c/\sqrt{2}$. It orders ferromagnetically below 160K[7] and has a saturation moment of $1.6\mu_B$ which is so far the largest known in any 4*d* ferromagnet[8]. $SrRuO_3$ is thus close to a half metal though the spin polarisation at the Fermi level has been predicted from band structure calculations to be in the range P=0.091-0.2[9,10]. The Fermi velocity of the minority spin carrier band has however been predicted to be 2-3 times larger than the majority band. This should give rise to a large negative transport spin polarisation. In the family of oxide ferromagnets with the perovskite crystal structure, the main attraction of $SrRuO_3$ stems from the fact that it is a "clean" system without any substitutional disorder. This, combined with the ability to grow very high quality single crystalline films, makes it possible to realise large mean free paths of the order of 500Å in clean samples[11]. The clean ballistic limit can therefore be reached in a point contact experiment. This limit is often difficult to attain in systems like hole doped rare-earth manganites where substitutional disorder severely limits the mean free path in the sample.

In this paper we report the transport spin polarisation in $SrRuO_3$ from Andreev reflection[12] data using a Nb point contact in the ballistic limit. The two films used in this study were grown in a way similar to the films used earlier for quantum oscillations studies[13]. The thicknesses of the films were ~2000Å. Conductance versus voltage (G-V) characteristics of the Nb-$SrRuO_3$ point contact were measured in the temperature range 2.6K-4.2K by dipping the sample and the tip in liquid He and pumping over the He bath. The Nb tip was fabricated either through electrochemical etching of Nb wire in potassium hydroxide solution or by mechanically polishing the tip. No significant difference was observed between the two different kind of tips. The conductance versus voltage of the point contact was measured directly using a modulation technique and averaged over ten sweeps for every (G-V) curve.

The residual resistivities of the two films used in this study were 6.8μΩ–cm and 5.6μΩ–cm for S1 and S2 respectively. Estimating the electronic mean free path ($l$) corresponding to these values is not trivial, but evidence from previous studies[11,13] suggests values of several hundred ångstroms. The diameter of the point contact ($d$) can be estimated from the normal state resistance ($R_N$) of the point contact using the approximate formula given by Wexler[14],

$$R_N \approx \frac{4}{3\pi}\frac{\rho l}{d^2} \quad \frac{\rho}{2d} \quad , \tag{3}$$

where $\rho$ is the resistivity of the sample. For the cleanest point contacts on both S1 and S2, $R_N$~12Ω. This corresponds to $d$~100Å. Thus we expect our measurements of transport spin polarisation to be well in the ballistic limit ($d<l$) of the point contact. An additional point to note is that the absolute resistance of both the films below 10K was ~0.5Ω. Correspondingly only point contacts with point contact resistance larger than 10Ω were analysed, to avoid any significant voltage contribution to the point contact spectra from the voltage drop in the sample.

In figure 1(a-f) we show some representative plots of the normalised conductance versus voltage of the Nb-$SrRuO_3$ point contact taken at 4.2K. The different spectra were recorded by engaging the point contact several times on the films. These different point contacts differ in the value of the scattering barrier at the interface between the ferromagnet and the superconductor, which depends on the microscopic details of the interface. To analyse the G-V characteristic of the point contact we use a modified Blonder-Tinkham-Klapwijk (BTK)[15] scheme, where the total current is decomposed into an unpolarised component and a fully polarised component[16,17]. There has been some controversy in the existing literature regarding the exact form of the reflection and transmission coefficients to be used for the analysis[18]. We are of the opinion that the approach of ref. 16 is the correct one and use it in our work. For the sake of completeness we briefly outline the model here. We consider the total current across our point contact to be made of an unpolarised component and a fully polarised component. In the model developed by BTK the current (I) is given terms of the Andreev reflection probability, A(E), and the normal reflection probability, B(E) of an incident electron on the ferromagnet superconductor interface as,

$$I \quad N v_F \int_{\infty}^{\infty} [f(E \quad eV,T) \quad f(E,T)][1 \quad A(E) \quad B(E)]\, dE = N v_F I' \quad . \tag{1}$$

Here N is the density of states in the ferromagnet, $v_F$ is the Fermi velocity and the coefficients A(E) and B(E) are different for the unpolarised current and the fully polarised current. Since in a typical Andreev spectrum the normalised conductance $G(V)/G_N$ ( $\equiv(dI/dV)/(dI/dV)_{eV>>\Delta}$) as a function of voltage is fitted it is enough to evaluate the integral $I'$ without considering the prefactor $Nv_F$ explicitly. For the unpolarised case BTK solved the Bogoliubov-de Gennes (BdG)[11] equations at the interface to find the coefficients A(E) and B(E). An incident particle on the interface (at $x$=0) given by $\Psi_{inc}=\begin{pmatrix}1\\0\end{pmatrix}e^{ikx}$ produces a reflected component $\Psi_{refl}=b\begin{pmatrix}e^{ikx}\\0\end{pmatrix}\;\; a\begin{pmatrix}0\\e^{ikx}\end{pmatrix}$ and a transmitted component $\Psi_{trans}=c\begin{pmatrix}u^S\\v^S\end{pmatrix}e^{iq_u x}\;\; d\begin{pmatrix}v^S\\u^S\end{pmatrix}e^{iq_v x}$. Here $u^S$ and $v^S$ are obtained from the solution of the BdG equation in the superconductor and are given by $(u^S)^2=1-(v^S)^2=(1/2)[1+\{(E^2-\Delta^2)/E^2\}^{1/2}]$. The first and second terms in $\Psi_{refl}$ correspond to the normal and Andreev reflection processes respectively. The coefficients $a$, $b$, $c$ and $d$ are calculated from the boundary conditions:

(i) $\Psi_n(x=0)=\Psi_s(x=0)$

(ii) $\Psi_s'(x=0)\;\;\Psi_n'(x=0)=\frac{2mV_0}{\hbar^2}\Psi(x=0)$ ,

where $\Psi_n(x)=\Psi_{inc}(x)\;\;\Psi_{refl}(x)$ , $\Psi_s(x)=\Psi_{tran}(x)$ are the wavefunctions inside the normal metal and the superconductor respectively. The interfacial scattering at the ferromagnet/superconductor interface is simulated through a delta function potential of the form $V_s(x)=V_0\delta(x)$ at the interface. $V_s(x)$ originates both from interfacial scattering from an imperfect junction (such as an oxide barrier on the metal electrode) and from the mismatch in the Fermi energies between the metal and the superconductor[19]. The coefficients A(E) and B(E) correspond to the probability currents associated with the Andreev and normal reflection process and are given by $A(E)=a^*a$ and $B(E)=b^*b$. In the fully polarised scenario, the allowed $k$ vectors are only in one spin direction and the Andreev reflected hole cannot propagate since it has a spin opposite to the incident electron. Therefore the Andreev reflected component gives rise to an evanescent wave. In this case the reflected component is given by $\Psi_{refl}=b\begin{pmatrix}e^{ikx}\\0\end{pmatrix}\;\; a\begin{pmatrix}0\\e^{\kappa x}\end{pmatrix}$ where κ is inversely proportional to the length over which the evanescent wave decays. The evanescent wave does not carry any current so A(E)=0. $B(E)=b^*b$ can be calculated using the same boundary conditions as before. In table I we list the A(E) and B(E) for the polarised and unpolarised case assuming $\kappa\rightarrow\infty$ [16]. For an arbitrary transport polarisation $P_t$ the total current will be given by,

$$I=I_u(1-P_t)+I_p P_t \qquad (2)$$

where $I_u$ and $I_p$ are given by equation (1) using the unpolarised and the polarised A(E) and B(E) coefficients respectively. The experimental point contact spectra are fitted (solid line in Fig. 1(a-f)) with the strength of the scattering barrier, Z (= $V_0/\hbar v_F$ ), the transport spin polarisation, $P_t$, and the superconducting energy gap, $\Delta$, as fitting parameters[20]. We believe that the difference between the fitted values of $\Delta$ for the point contacts on S1 and S2 arises from variation in the quality of tips produced by our tip fabrication process. Note, however, the consistency of the value fitted from different spectra using the same tip, and that (as expected) none of the values exceed that of bulk Nb. The value of $P_t$ extracted from the fits decreases with increasing Z, the scattering strength at the interface. This behaviour has been observed earlier in iron, cobalt and nickel films as well as $CrO_2$ and in manganites[17,21]. It is believed to be due to the spin mixing effect at the magnetically disordered scattering barrier formed at the interfaces[22,23]. In that sense Andreev reflection provides a lower bound on the transport spin polarisation.

In order to extract the intrinsic spin polarisation we plot (in figure 2) $P_t$ as a function of Z obtained from the Andreev spectra at 4.21K from both samples S1 and S2. Within experimental errors, we could fit our Z dependence of $P_t$ with a parabolic curve for datapoints obtained from both S1 and S2. Thus we extract $P_t$ in the limit $Z\rightarrow 0$ by extrapolating back a fitted parabolic curve. The intrinsic polarisation obtained from the fit is $P_t=0.51\pm0.02$.

The fact that datapoints obtained from both S1 and S2 can be fitted with a single parabolic curve giving the same value for $P_t$ though their residual resistances are different is important. Mazin *et al.*[16] have pointed out that in the diffusive limit of a point contact ($d>>l$) the transport spin polarisation is given by $P_t=(\langle N_\uparrow v_{F\uparrow}^2\rangle-\langle N_\downarrow v_{F\downarrow}^2\rangle)/(\langle N_\uparrow v_{F\uparrow}^2\rangle+\langle N_\downarrow v_{F\downarrow}^2\rangle))$ instead of $P_t=(\langle N_\uparrow v_{F\uparrow}\rangle-\langle N_\downarrow v_{F\downarrow}\rangle)/(\langle N_\uparrow v_{F\uparrow}\rangle+\langle N_\downarrow v_{F\downarrow}\rangle))$. Since these two quantities are normally different for a ferromagnet in a borderline case, when $d\approx l$, one can see a systematic change in the measured value of $P_t$, as a function of sample disorder[24]. The unique value of the spin polarisation obtained for both the films further confirms that our measurements are well in the ballistic limit ($d<<l$) of the point contact.

As a further check on our experiment and data analysis procedure we studied the temperature dependence for the point contact with lowest Z on S1. As expected, the extracted values of $P_t$ ($=0.48\pm0.02$) extracted from the spectra are constant within error bars over this temperature range.

It is interesting to note that the value of $P_t$ measured in this study is much larger than the spin polarisation P predicted from band structure calculations[9,10]. Though the two band structure calculations so far published on this compound differ in their detail the predicted value of P is small: P=0.091[10] and P=0.2[9] respectively. The large value of $P_t$ compared to the value of spin polarisation P predicted from band structure calculations suggests that the large transport spin polarisation originates primarily from a difference in the Fermi velocities of minority and majority spin bands[25]. This is also in agreement with both band structure calculations which predict the average Fermi velocity of the carriers in the majority spin band to be smaller by a factor of 2-3 than that in the minority spin band.

In this context it is also interesting to compare our results with the spin polarisation measured by Worledge and Geballe using the Meservey-Tedrow technique[26]. In the Meservey Tedrow technique the spin polarisation is measured by fabricating a tunnel junction on the ferromagnet with a superconducting counter-electrode and studying its conductance spectra with and without an applied magnetic field. Within a simple model the spin polarisation measured with this technique is similar to that given by Andreev reflection in the diffusive limit. The main advantage of this technique over Andreev reflection is that it is also sensitive to the sign of the spin polarisation. Consistent with band structure predictions Worledge and Geballe observed a negative transport spin polarisation in $SrRuO_3$.. However, the degree of spin polarisation obtained by them (~9.5%) is much smaller than the transport spin polarisation expected from band structure calculations. The main disadvantage of the Meservey-Tedrow technique is that the measured transport spin polarisation depends on the spin decay length in the tunnel barrier[26]. Therefore the degree of spin polarisation measured with this technique depends on the insulating spacer material used, and may not reflect the true spin polarisation in the material. Andreev reflection in the clean limit on the other hand does not suffer from this drawback.

Finally, we would like to note that most experiments involving the injection of spins in a *d*-wave superconductors have used doped rare-earth manganites such as $La_{0.7}Sr_{0.3}MnO_3$ or $La_{0.7}Ca_{0.3}MnO_3$ as the natural choice, in view of their half-metallic character observed from spin polarised photoemission experiments[27]. However, in recent times several experiments have cast doubt on the use of this material as the ideal choice for spin injection. Point contact Andreev reflection measurements by Nadgorny *et al.*[24] on several samples of $La_{0.7}Sr_{0.3}MnO_3$ with varying level of non-subtitutional disorder showed the polarisation to be much less than 100%. Measurements of spin polarisation by Ji *et al.*[28] using the same technique in the diffusive limit on single crystals of $La_{0.7}Sr_{0.3}MnO_3$ and $La_{0.6}Sr_{0.4}MnO_3$ support these results. In addition, spin polarised photoemission also showed that the spin polarisation of the surface layer decays much more rapidly with increasing temperature than the bulk[29]. It is therefore interesting to note that the transport spin polarisation of $SrRuO_3$ in the ballistic limit is comparable to $La_{0.7}Sr_{0.3}MnO_3$. However, the advantage of $SrRuO_3$ is that high quality thin films of this material have much lower residual resistance owing to the absence of substitutional disorder and the ability to grow very high quality single crystalline films, thereby reducing the problems associated with Joule heating in spin injection experiments. It would therefore be interesting to compare the effect of spin injection from $SrRuO_3$ with the data from doped manganites.

In summary, we have measured the transport spin polarisation in epitaxial thin films of $SrRuO_3$ in the ballistic limit. The transport spin polarisation in this compound is comparable to the transport spin polarisation measured in doped rare-earth manganites. Comparing with the value of the spin polarisation (P) calculated from band structure predictions, we conclude that the large transport spin polarisation is mainly due to the difference in the Fermi velocities of the majority and minority spin carrier electrons. This agrees well with the prediction from band structure calculations.

*Acknowledgement:* We would like to thank the Leverhulme Trust and the Royal Society for research grants supporting this work and G. R. Walsh and D. A. Brewster for help with constructing the experimental apparatus and S. A. Grigera for critically reading the manuscript. We are also grateful to M.S. Osofsky for sharing his group's results with us prior to publication.

**Table 1***: ( $\gamma^2 = (((u^S)^2 \quad (v^S)^2) Z^2 \quad (u^s)^2)^2$ , $Z = V_0 / \hbar v_F$ )

| | **E<Δ** | **E>Δ** |
|---|---|---|
| **$A_u$(E)** | $\frac{\Delta^2}{E^2 \quad (\Delta^2 \quad E^2)(1 \quad 2Z^2)^2}$ | $\frac{(u^S v^S)^2}{\gamma^2}$ |
| **$B_u$(E)** | 1-A(E) | $\frac{((u^S)^2) \quad ((v^S)^2)^2 Z^2 (1 \quad Z^2)}{\gamma^2}$ |
| **$A_p$(E)** | 0 | 0 |
| **$B_p$(E)** | 1 | $\frac{\left(\left(\frac{E^2 \quad \Delta^2}{E^2}\right)^{1/2} \quad 1\right)^2 \quad 4Z^2\left(\frac{E^2 \quad \Delta^2}{E^2}\right)}{\left(\left(\frac{E^2 \quad \Delta^2}{E^2}\right)^{1/2} \quad 1\right)^2 \quad 4Z^2\left(\frac{E^2 \quad \Delta^2}{E^2}\right)}$ |

*The subscripts u and p denote the coefficients for the unpolarised and the polarised current respectively.

## Figure Captions

**Figure 1.** (a-c) Representative $G(V)/G_N$ versus V plots for the $SrRuO_3$/Nb point contact measured at 4.2K for S1; (d-f) the same for S2. The solid lines are the fits to the BTK model. The measured value of $P_t$ decreases from the intrinsic value for dirtier contacts, i.e. larger Z.

**Figure 2.** $P_t$ as a function of Z for different point contacts. The triangles and diamonds on the $P_t$ versus Z plot are data points taken from samples S1 and S2 respectively. The dashed line is a parabolic fit to the data to extract the transport spin polarisation for Z=0. The inset shows the transport spin polarisation for S1 measured at different temperatures. The solid line is a guide to the eye.

## Table Caption

**Table 1.** The coefficients A(E) and B(E) corresponding to the polarised and unpolarised case.